\documentclass[%
 reprint,
superscriptaddress,
 amsmath,amssymb,
 aps,
]{revtex4-2}

\usepackage{graphicx}% Include figure files
\usepackage{dcolumn}% Align table columns on decimal point
\usepackage{bm}% bold math
\usepackage{threeparttable,booktabs}
\begin{document}

\preprint{APS/123-QED}

\title{First application of mass measurement with the Rare-RI Ring reveals the solar $r$-process abundance trend at $A=122$ and $A=123$} %:\\with Forced Linebreak}% Force line breaks with \\

\author{H. F. Li}
 
\affiliation{Institute of Modern Physics, Chinese Academy of Sciences, Lanzhou 730000, People’s Republic of China}
\affiliation{Lanzhou University, Lanzhou 730000, People’s Republic of China}
\affiliation{RIKEN Nishina Center, RIKEN, Saitama 351-0198, Japan}
\affiliation{University of Chinese Academy of Sciences, Beijing 100049, People’s Republic of China}

\author{S. Naimi}
\email{Corresponding author: snaimi(at)ribf.riken.jp}
\affiliation{RIKEN Nishina Center, RIKEN, Saitama 351-0198, Japan}%

\author{T.~M.~Sprouse}
\author{M.~R.~Mumpower}
\affiliation{Theoretical Division, Los Alamos National Laboratory, Los Alamos, New Mexico, 87545, USA}%

\author{Y. Abe}
\author{Y. Yamaguchi}
\author{D. Nagae}
 \altaffiliation[Current affiliation: ]{Research Center for SuperHeavy Elements, Kyushu University, Fukuoka, Fukuoka 819-0395, Japan}
 \author{F. Suzaki }
  \altaffiliation[Current affiliation: ]{Advanced Science Research Center, Japan Atomic Energy Agency, Ibaraki 319-1195, Japan}
  \author{M. Wakasugi}
\affiliation{RIKEN Nishina Center, RIKEN, Saitama 351-0198, Japan}%

\author{H. Arakawa}
\author{W.B. Dou}
\author{D. Hamakawa}
\author{S. Hosoi}
\author{Y. Inada}
\author{D. Kajiki}
\author{T. Kobayashi}
\author{M. Sakaue}
\author{Y. Yokoda}
\author{T. Yamaguchi}
\affiliation{Department of Physics, Saitama University, Saitama 338-8570, Japan}

\author{R. Kagesawa}
\author{D. Kamioka}
\author{T. Moriguchi}
\author{M. Mukai}
 \altaffiliation[Current affiliation: ]{RIKEN Nishina Center, RIKEN, Saitama 351-0198, Japan}
\author{A. Ozawa}
\affiliation{Institute of Physics, University of Tsukuba, Ibaraki 305-8571, Japan}

\author{S. Ota}
\altaffiliation[Current affiliation: ]{Research Center for Nuclear Physics, Osaka University, Osaka, 567-0047, Japan}
\author{N. Kitamura}
\author{S. Masuoka}
\author{S. Michimasa}
\affiliation{Center for Nuclear Study, University of Tokyo, Wako, Saitama 351-0198, Japan}

\author{H. Baba}
\author{N. Fukuda}
\author{Y. Shimizu}
\author{H. Suzuki}
\author{H. Takeda}
\affiliation{RIKEN Nishina Center, RIKEN, Saitama 351-0198, Japan}%

\author{D.S. Ahn}
\affiliation{RIKEN Nishina Center, RIKEN, Saitama 351-0198, Japan}%
\affiliation{Center for Exotic Nuclear Studies, Institute for Basic Science (IBS),
Daejeon 34126, Republic of Korea}%

\author{M. Wang}
\author{C.Y. Fu}
\author{Q. Wang} 
\author{S. Suzuki}
\author{Z. Ge}
 \altaffiliation[Current affiliation: ]{GSI Helmholtzzentrum f{\"u}r Schwerionenforschung, Planckstra$\ss$e 1, 64291 Darmstadt, Germany}
\affiliation{Institute of Modern Physics, Chinese Academy of Sciences, Lanzhou 730000, People’s Republic of China}

\author{Yu.~A. Litvinov}
\affiliation{GSI Helmholtzzentrum f{\"u}r Schwerionenforschung, Planckstra$\ss$e 1, 64291 Darmstadt, Germany}

\author{G. Lorusso}
\affiliation{National Physical Laboratory, Teddington, TW11 0LW, United Kingdom}
\affiliation{Department of Physics, University of Surrey, Guildford GU2 7XH, United Kingdom}
\author{P.~M. Walker}
\author{Zs. Podolyak }
\affiliation{Department of Physics, University of Surrey, Guildford GU2 7XH, United Kingdom}

\author{T. Uesaka}
\affiliation{RIKEN Nishina Center, RIKEN, Saitama 351-0198, Japan}%

\date{\today}% It is always \today, today,

\begin{abstract}
The Rare-RI Ring (R3) is a recently commissioned cyclotron-like storage ring mass spectrometer dedicated to mass measurements of exotic nuclei far from stability at Radioactive Isotope Beam Factory (RIBF) in RIKEN. 
The first application of mass measurement using the R3 mass spectrometer at RIBF is reported. 
Rare isotopes produced at RIBF, $^{127}$Sn, $^{126}$In, $^{125}$Cd, $^{124}$Ag, $^{123}$Pd, were injected in R3. 
Masses of $^{126}$In, $^{125}$Cd, and $^{123}$Pd were measured whereby the mass uncertainty of $^{123}$Pd was improved. 
This is the first reported measurement with a new storage ring mass spectrometery technique realized at a heavy-ion cyclotron and employing individual injection of the pre-identified rare nuclei.
The latter is essential for the future mass measurements of the rarest isotopes produced at RIBF.
The impact of the new $^{123}$Pd result on the solar $r$-process abundances in a neutron star merger event is investigated by performing reaction network calculations of 20 trajectories with varying electron fraction $Y_e$. 
It is found that the neutron capture cross section on $^{123}$Pd increases by a factor of 2.2 and $\beta$-delayed neutron emission probability, $P_\mathrm{1n}$, of  $^{123}$Rh  increases by 14\%. 
The neutron capture cross section on $^{122}$Pd decreases by a  factor of 2.6 leading to pileup of material at $A=122$, thus reproducing the trend of the solar $r$-process abundances. 
The trend of the two-neutron separation energies (S$_\mathrm{2n}$) was investigated for the Pd isotopic chain. 
The new mass measurement with improved uncertainty excludes large changes of the S$_\mathrm{2n}$ value at $N=77$. 
Such large increase of the S$_\mathrm{2n}$ values before $N=82$ was proposed as an alternative to the quenching of the $N=82$ shell gap to reproduce $r$-process abundances in the mass region of $A=112-124$.
\end{abstract}

\maketitle

%\tableofcontents

%%%%%%%%%%%%%%%%%
% Introduction
%%%%%%%%%%%%%%%%%
\begin{figure*}[hbt]
\includegraphics[width=18cm]{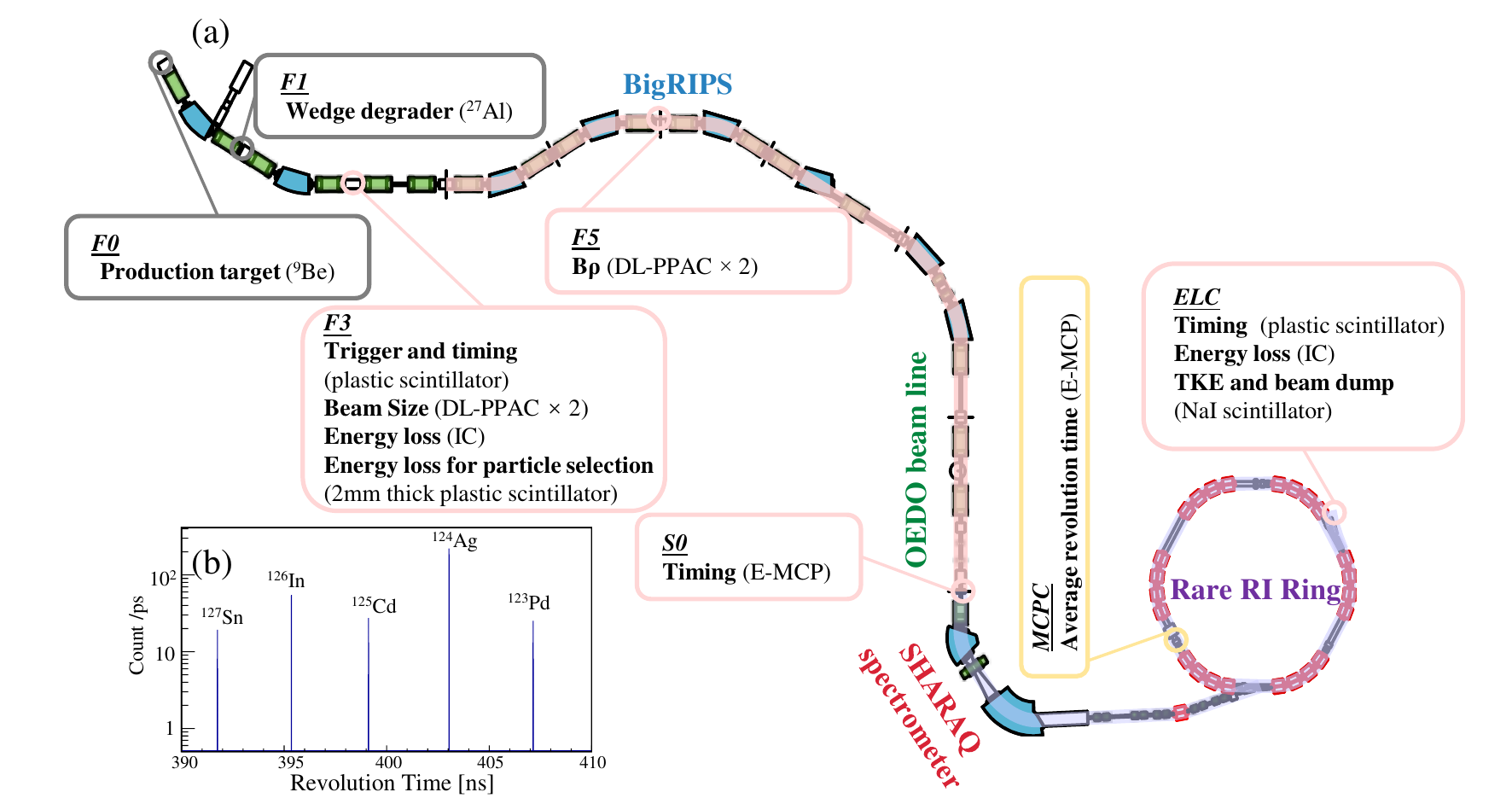}
\caption{\label{fig:setup} (a) Configuration of the detectors installed in the beam line and Rare RI Ring (R3). (b) The revolution time spectrum for the nuclei measured in this work. The background free spectrum reflects the efficiency of event-by-event tracking.   
}
\end{figure*}

The discovery of the historical GW170817 event of binary neutron stars merger and the subsequent kilonova AT2017go \cite{Abbott_2017} for the GW170817 \cite{Abbott2017gw170817} was a major milestone toward revealing the secret of the synthesis of heavy elements via the rapid neutron capture process ($r$-process)\cite{Burbidge1957}.
The recent identification of strontium in the kilonova radiation gave a strong evidence of the production of $r$-process elements \cite{Watson2019}.
However, modeling of the accretion disk formed in supernova-triggered collapse of rapidly rotating massive stars or collapsars, showed that $r$-process elements could be also produced in considerable amounts \cite{Siegel2019}.
The presence of $r$-process heavy elements was also observed in the dwarf galaxy Reticulum II \cite{Ji2016}, where the accretion disk of collapsars might be the main source of production. 
Heavy elements might be therefore synthesized in various astrophysical sites such as neutron star mergers and magneto-rotational supernovae \cite{Cowan2021,Horowitz_2019}.
To model the formation of heavy chemical elements under different astrophysical conditions, a large and diverse amount of nuclear data is needed, especially for neutron-rich nuclei that live for a fraction of a second.  
Nuclear masses are  important ingredients since they reflect the neutron separation energies, which are required for the determination of neutron capture and photodissociation rates \cite{Burbidge1957,Surman2009,Mumpower2015}. 
A vast number of neutron-rich nuclei involved in the $r$-process can now be produced in the laboratory at rare isotope facilities and their properties measured with high precision. 
However, all nuclei needed for modelling the $r$-process will not be accessible even at the new-generation radioactive-ion beam facilities. 
A robust model based on accurate properties of neutron-rich nuclei is thus essential to reveal the astrophysical conditions in which heavy elements could be produced. 
Such a model will help quantify the production rates in various sites, which will result in more accurate nucleosynthesis calculations capable of reproducing the $r$-process elements' chemical abundances \cite{Cote2017,Hotokezaka2018}.

This Letter reports precision mass measurements of neutron-rich nuclei produced at the Radioactive Isotope Beam Factory (RIBF) and their implication in the production of $r$-process elements with atomic mass number $A=122$ and $A=123$. 
Mass measurements of nuclei with neutron number $N=77$ were performed for the first time with a new type of mass spectrometer, namely the Rare-RI Ring (R3), recently commissioned at the RIBF/RIKEN facility \cite{Nagaestori17}. 
We examine the implication of the $^{123}$Pd mass on the abundance calculation for conditions representing a neutron star merger event. 
These first mass measurements at RIBF of neutron-rich isotopes in a remote region of the nuclear chart open a door to reaching $r$-process nuclei at $N=8$2 and beyond.

%%%%%%%%%%%%%%%%%%%%%%%%%%%%%%%%%%%%%%%
% Experiment description
%%%%%%%%%%%%%%%%%%%%%%%%%%%%%%%%%%%%%%%

In the experiment, the secondary beam was produced by in-flight-fission of the 345~MeV/nucleon $^{238}$U beam provided by the Superconducting Ring Cyclotron (SRC) impinged on the 6~mm  thick beryllium target which was placed upstream of the BigRIPS separator at F0 focal plane (see Fig. \ref{fig:setup}).
The secondary fragments of interest were separated by the first stage of the BigRIPS as described in \cite{FUKUDA2013323}.
For this purpose, a 5~mm wedge-shaped degrader was introduced at the F1 focal plane of the BigRIPS.
The magnetic rigidity B$\rho$ and the transmission efficiency were optimized for the reference particle $^{124}$Ag.
The momentum selection was done by setting the slits at F1 to $\pm$2~mm, corresponding to the R3 momentum acceptance of $\pm$0.3\%.
The injection kicker magnets system placed inside the R3 is limited to a repetition rate of 100~Hz. 
Therefore, to accept the quasi-continuous beam from the SRC, the individual self-injected trigger technique was developed for injecting pre-identified particles of interest \cite{2021YYamaguchi}.
The particle identification (PID) was achieved by the $\Delta$E-TOF method in the beam line, where $\Delta$E is the energy loss measured by the ionization chamber (IC) placed at F3 and TOF is the time-of-flight measured by the plastic scintillator at F3 and the E-MCP detector \cite{NAGAE2021} at S0 of the SHARAQ spectrometer.
Also a 2-mm thick plastic scintillator was placed after the IC at F3 to get a rough $\Delta$E information needed for removing contaminants \cite{Abe2019}.
Two position monitors PPACs (Parallel Plate Avalanche Counter) were installed at F3 to monitor the beam size and two double PPACs were installed at F5 which is a dispersive focal plane to measure B$\rho$ of every individual particle prior to its injection into the R3.
The particle circulated in the R3 for about 1800 revolutions before it was ejected from the ring.
The total TOF in the R3 was measured by the E-MCP detector at S0 and a plastic scintillator detector placed at ELC after the ejection from the R3.
Another IC was installed at ELC, where an additional PID was performed. 
Finally, particles were stopped in the NaI scintillator detector placed behind the IC at ELC.

\begin{figure}[!ht]
\includegraphics[width=\columnwidth]{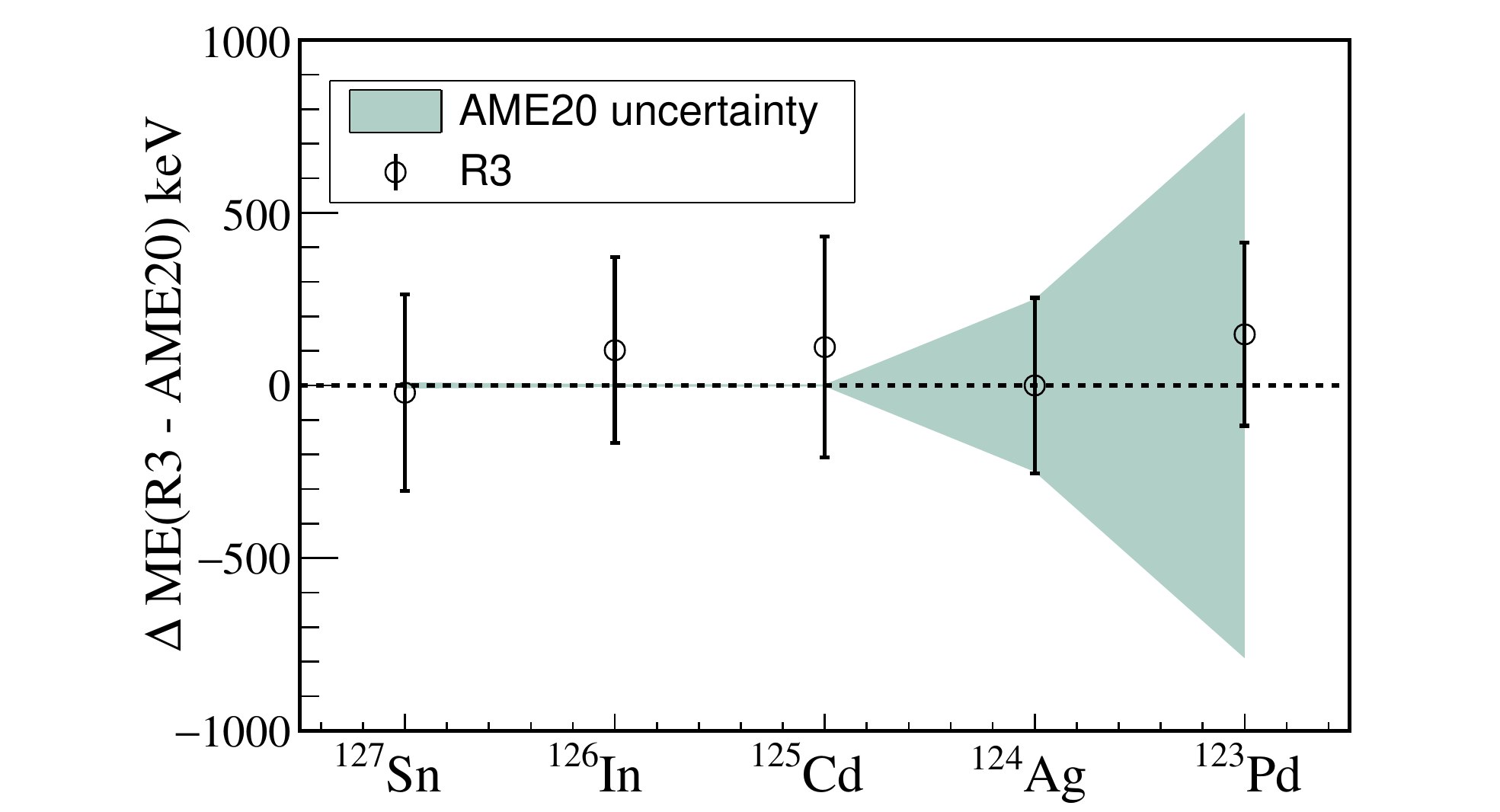}
\caption{\label{fig:mass} Mass excess values of nuclei measured at R3 compared to literature values from AME2020 \cite{Wang_2021}}
\end{figure}

The mass-to-charge ratio ($m/q$) of the particle of interest with a revolution time $T$ is determined relative to a reference particle with m$_{0}$/q$_{0}$ and $T_0$ by using the following formula \cite{2012Ozawa,Nagaestori17}: 
\begin{equation}
\frac{m}{q} = \frac{m_0}{q_0}\frac{T}{T_0}\sqrt{\frac{1-\beta^2}{1-\left(\frac{T}{T_0}\beta^2\right)}}, \label{r3_prin}
\end{equation}
where $\beta$ is the velocity of the particle of interest relative to the speed of light in vacuum. 
Revolution time spectrum of all injected nuclei is shown in Fig. \ref{fig:setup} (b) (details of determination of the revolution time in R3 can be found in \cite{Naimi_2020}). 
Since the isochronous condition of the ring is optimized for the reference particle, $T_0$ is independent of the momentum. 
To determine the mass, the velocity $\beta$ needs to be determined event-by-event from the time-of-flight along the beamline from F3 to S0 ($TOF_{3S0}$) by using the following equation,
\begin{equation}
\beta = \frac{Length_{3S0}}{\left(TOF_{3S0}+TOF_{offset}\right)}. \label{eq_beta}
\end{equation}
The average path length from F3 to S0 ($Length_{3S0}$) and the TOF$_{offset}$ caused by the electronics and the energy loss in the detectors on the beamline, are determined via Eq.(\ref{eq_beta}) by using known masses of $^{124}$Ag and  $^{127}$Sn. 
The parameters that could reproduce the known $m/q$ values are $Length_{3S0} = 84.859(2)$~m and $TOF_{offset} = 325.47(1)$~ns.  
The mass is then determined for each event via Eq.(\ref{r3_prin}).
Additional systematic uncertainties, $\sigma_{sys}$, due to the determination of parameters such as $Length_{3S0}$, $TOF_{offset}$ and $T_0$ were estimated and reported in Table \ref{tab:mass}. 
Details of data analysis method can be found in references \cite{Naimi_2020,Nagaestori17}.
The full data analysis method as well as the details of estimating the systematic uncertainties will be reported in a subsequent publication.  
The determined mass excess values are listed in Table \ref{tab:mass}. 
Comparison with literature values from the latest Atomic Mass Evaluation, AME2020 \cite{Wang_2021}, are plotted in Fig. \ref{fig:mass}.
As shown in Table \ref{tab:mass}, the uncertainties are dominated by the mass uncertainty of the reference particle $^{124}$Ag. 
The choice of this reference instead of $^{125}$Cd, which has lower uncertainty, is mainly due to the presence of a long-lived isomeric state at 186~keV in the latter that is difficult to separate with R3. 
The mass precision was therefore scarified for higher accuracy. 
However, if the mass of $^{124}$Ag is remeasured with higher precision, the uncertainties of all other masses will be reduced. 
\begin{table}[h]
\caption{\label{tab:mass}Mass excess values from literature and measured in this work are shown in the second and third column, respectively. 
Total uncertainties are shown as well as the contribution from the reference mass uncertainty $\sigma_{m_0}$ and the statistical uncertainty $\sigma_{stat}$. 
The systematic uncertainty $\sigma_{sys}$ is estimated from the uncertainty of $T_0$ and the fit parameters $Length_{3S0}$ and $TOF_{offset}$ in Eq.(\ref{eq_beta}).}
\centering
\begin{tabular}{@{}lccccccc@{}}
\toprule
Nucleus & ME$_{AME20}$ & ME$_{R3}$ &$\sigma_{total}$  & $\sigma_{m_0}$ & $\sigma_{stat}$ & $\sigma_{sys}$  \\
&  [keV] &[keV] &[keV] &[keV] &[keV] & [keV] \\
\midrule
$^{126}$In&   -77809(4)	      &-77707   &269 & 254 & 65  & 62 \\
$^{125}$Cd&   -73348.1(29)    &-73237   &320 & 252 & 192 & 40 \\
$^{123}$Pd&   -60430(790)     &-60282   &265 & 248 & 86  & 40 \\
\bottomrule
\end{tabular}
\end{table}

\begin{figure}[!ht]
\includegraphics[width=\columnwidth]{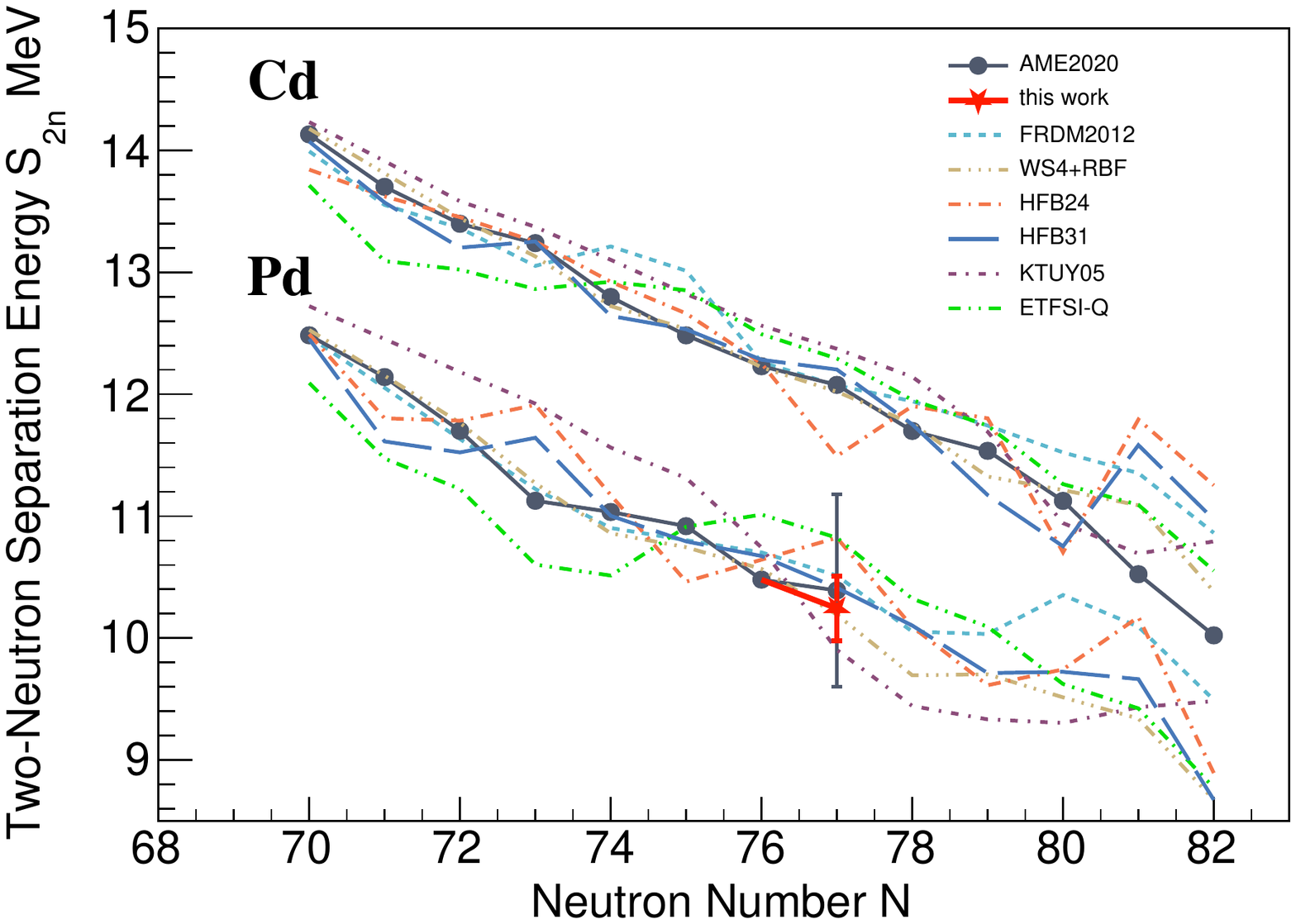}
\caption{\label{fig:s2n} The two neutron separation energy values (S$_\mathrm{2n}$) for Cd and Pd isotopic chains. Experimental data are taken from AME2020 \cite{Wang_2021}. The S$_\mathrm{2n}$ value derived from our new mass value of $^{123}$Pd is shown in red, while the theoretical predictions are shown  by the colored dash lines \cite{Moller2016,wang2014surface,goriely2013further,goriely2016further,Kura2005,pearson1996nuclear}}
\end{figure}

%%%%%%%%%%%%%%%%%%%%%%%%%%%%%%%%%%%%%%%
% Nuclear physics part
%%%%%%%%%%%%%%%%%%%%%%%%%%%%%%%%%%%%%%%
Subtle nuclear structure effects of neutron-rich nuclei are believed to have strong impact on the $r$-process abundances \cite{Grawe_2007,Arcones2011}. 
Failure to produce enough material for masses at A$=112-124$ was thought to be due to the shell quenching at $N=82$ \cite{PEARSON1996455}.
However, it has been recently demonstrated through mass measurement of $^{132}$Cd that this shell gap, although reduced, is not quenched for $Z<50$ \cite{Manea2020}.
Another suggested solution to this issue was the increase of the two-neutron separation energies (S$_\mathrm{2n}$) that might be associated with a sudden transition from deformed shape to spherical shape for nuclei with $Z<50$ and $N=75-82$ \cite{Grawe_2007}. 
In Fig. \ref{fig:s2n}, the S$_\mathrm{2n}$ values for Cd and Pd isotopic chains are shown with the updated value for the most neutron-rich Pd isotope at $N=77$. 
Global mass models (FRDM \cite{Moller2016}, KTUY05 \cite{Kura2005}, EFTSI-Q \cite{pearson1996nuclear} and WS4+RBF \cite{wang2014surface}) are also plotted in addition to the microscopic mass model HFB24 \cite{goriely2013further} and its more recent version HFB31 \cite{goriely2016further}. 
Changes in the S$_\mathrm{2n}$ values of the most recent mass models are significantly less pronounced for Pd isotopes as compared to earlier mass models such as ETFSI-Q, KTUY05 and HFB24. 
The new S$_\mathrm{2n}$ value of $^{123}$Pd shows a smooth decrease following the trend of the mass surface.  
Furthermore, the improvement of the uncertainty excludes significant large change in the S$_\mathrm{2n}$ value at $N=77$ as compared to the uncertainty of the $^{123}$Pd mass reported in the AME2020.  
%Based on the estimation of our systematic uncertainties, the mass uncertainty could be reduced to about 100~keV if the mass of $^{124}$Ag is remeasured with a precision of less than 30~keV. 

%%%%%%%%%%%%%%%%%%%%%%%%%%%%%%%%%%%%%%%
% Astrophysics part
%%%%%%%%%%%%%%%%%%%%%%%%%%%%%%%%%%%%%%%
We simulate the impact of the mass measurement of $^{123}$Pd in the $r$-process by employing the Portable Routines for Integrated nucleoSynthesis Modeling (PRISM) reaction network \cite{Sprouse2020, Sprouse2021}. 
The baseline nuclear physics properties are simulated with FRDM2012 \cite{Moller2015, Moller2016, Mumpower2016, Mumpower2018, Moller2019}. 
The mass of $^{123}$Pd in the baseline model is also taken from the FRDM2012. 
Changes to the mass propagate to cross sections and branching ratios in neighboring nuclei as in reference \cite{Mumpower2015}. 
We find that the changes in the capture cross sections, and $\beta$-delayed neutron probabilities (discussed below) have a significant effect as compared with propagation to separation energies alone.  
Past studies have indicated that the mass of $^{123}$Pd is most influential for abundances of nearby isobars~\cite{Mumpower2016r}, in particular the abundances of $A=122$ and $A=123,$ so we analyze the impact of the measured mass from this perspective.

\begin{figure}[htb]
\includegraphics[width=\columnwidth]{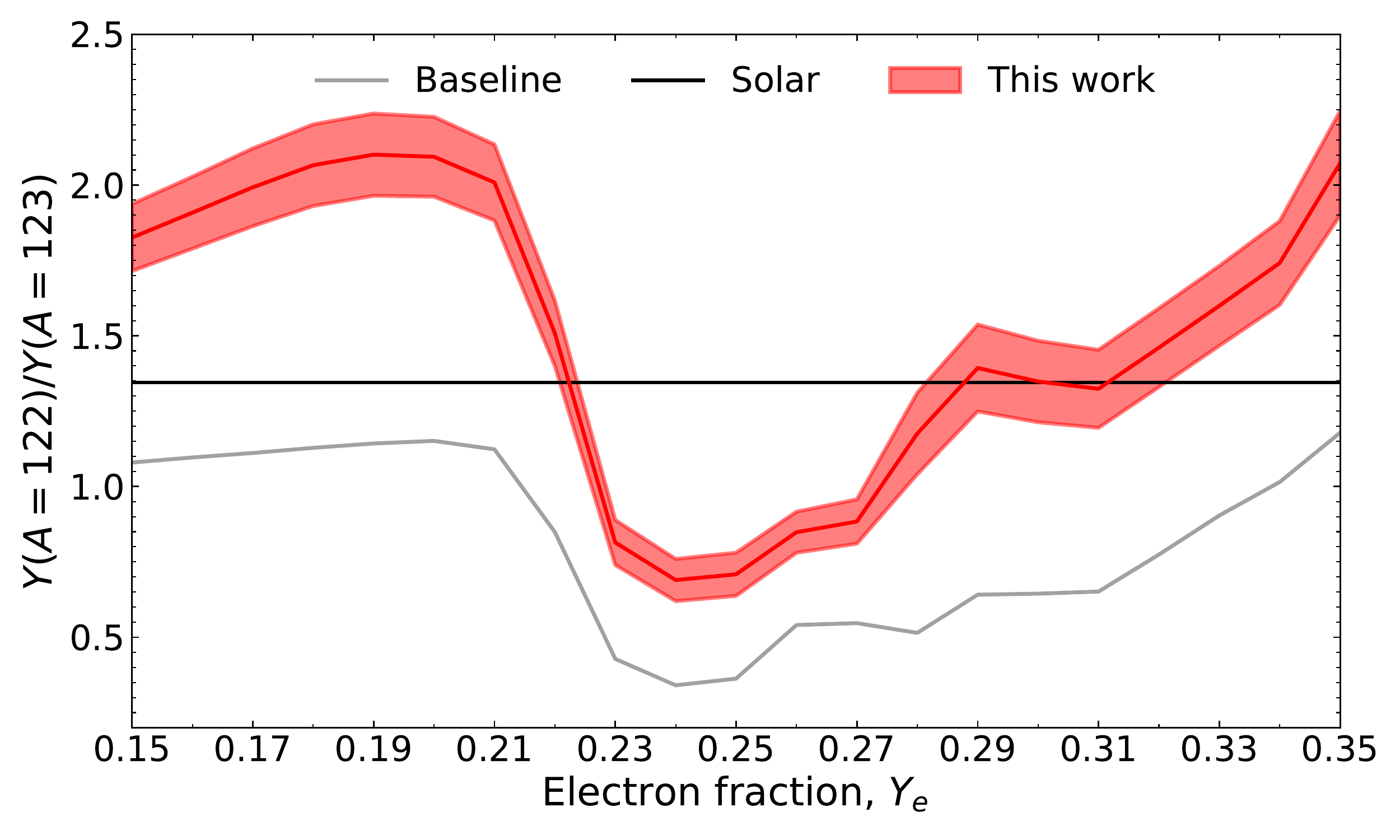}
\caption{\label{fig:rprocess_impact} Ratio of the $A=122$ to $A=123$ isotopic abundances as a function of electron fraction for the baseline model (gray line) and if our new mass measurement (red line) and its uncertainty (red band) are included. The horizontal black line indicates the value of the same ratio in the solar $r$-process residuals of Ref.~\cite{Arnould2007}. }
\end{figure}

Since there are uncertainties in the astrophysical conditions that could produce nuclei in the mass $A\sim120$ range, we simulate nucleosynthesis for a set of 20 parameterized $r$-process trajectories \cite{Zhu2018} with specific entropy $40\ k_B/\textrm{baryon}$, timescale $\tau=20\ \textrm{ms},$ and electron fraction running from $Y_e=0.15$ to $Y_e=0.35$ in intervals of $0.01$.
Figure \ref{fig:rprocess_impact} shows the change in the $A=122$ to $A=123$ abundance ratio for each trajectory due to the $^{123}$Pd mass measurement when compared against the baseline model.
For all trajectories considered here, the overall effect is the increase of the ratio with respect to the baseline, with the largest effects noted for the lowest electron fractions considered, $Y_e < 0.22$. 
In all of these trajectories, the baseline model fails to achieve the solar ratio, such that no combination of these trajectories would be able to reproduce this observed feature. 
However, when we update the nuclear data to include the measured $^{123}$Pd mass, the ratio is sufficiently varied across the range of trajectories, such that suitable linear combinations of these make reproduction of this ratio a viable possibility. 
This can be achieved, for example, by mixing contributions from trajectories that overproduce or underproduce $A=122$ nuclei relative to $A=123$ nuclei.
The effect arises due to changes in calculated nuclear properties introduced by the newly measured $^{123}$Pd mass. 
The neutron capture cross section for $^{122}$Pd decreases by a factor of 2.6 and for $^{123}$Pd increases by a factor of 2.2, while the $P_{1n}$ value, probability for the $\beta$-delayed neutron emission, of $^{123}$Rh increases by 14\% with the updated mass.
This results in an effective pileup of material along the $A=122$ isobar relative to the baseline. 
We note that sizeable statistical model uncertainties exist in the $\gamma$-strength function and level density of neutron-rich nuclei; here we consider the impact of the current measurement in isolation to other uncertainties. 

%%%%%%%%%%%%%%%%%%%%%%%%%%%%%%%%%%%%%%%
% End of astrophysics part
%%%%%%%%%%%%%%%%%%%%%%%%%%%%%%%%%%%%%%%

In summary, the first implementation of mass measurements performed by the Rare-RI Ring at the RIBF facility is reported. 
The most neutron-rich nuclei below the doubly magic nucleus $^{132}$Sn were studied, opening a new avenue to perform mass measurements of r-process nuclei near $N=82$. 
To reproduce the $r$-process abundances with masses $A=112-124$, an increase of S$_\mathrm{2n}$ values just before the $N=82$ for neutron-rich nuclei with $Z<50$ was proposed as an alternative to $N=82$ shell gap quenching. 
The new mass measurement of $^{123}$Pd and its uncertainty exclude large change of the S$_\mathrm{2n}$ value at $N=77$ for the Pd isotopic chain.
We performed calculations to estimate the impact of the $^{123}$Pd mass measured in the $r$-process. 
We found if our new mass value is used instead of the FRDM value, that the $r$-process abundances at $A=122$ and $A=123$ are modified towards being more consistent with solar values.  
This indicates that the $r$-process calculations are very sensitive to masses in this region since a change of $^{123}$Pd mass by just 478~keV causes a sizeable effect. 
This finding highlight the need for high precision mass measurements to address the $r$-process in this mass region.

\begin{acknowledgments}
We are grateful to the RIKEN RIBF accelerator crew and CNS, University of Tokyo for their efforts and supports to operate the RI beam factory. 
H.F. L. expresses gratitude to the RIKEN International Program Associate. 
This work was supported by the RIKEN Pioneering Project Funding  (``Extreme precisions to Explore fundamental physics with Exotic particles") and JSPS KAKENHI Grant Nos. 19K03901, 26287036, 25105506, 15H00830, 17H01123, 18H03695, 17K14311.
%Matt & Trevor
T.M.S. and M.R.M. were supported by the US Department of Energy through the Los Alamos National Laboratory (LANL). 
LANL is operated by Triad National Security, LLC, for the National Nuclear Security Administration of U.S.\ Department of Energy (Contract No.\ 89233218CNA000001).
T.M.S. was partly supported by the Fission In R-process Elements (FIRE) Topical Collaboration in Nuclear Theory, funded by the U.S. Department of Energy. 
%Yury
Yu.A. L. acknowledges support by the European Research Council (ERC) under the European Union's Horizon 2020 research and innovation program (grant agreement No 682841 ``ASTRUm'').
% Phil Walker
This work is supported by the UK Science and Technology Facilities Council under grant no. ST/P005314/1 and the National Natural Science Foundation of China (Grant No. 11975280).
%Zhuang  & Wang Qian
%This work is supported by the National Natural Science Foundation of China (Grant No. 11975280).
\end{acknowledgments}

\nocite{*}

\bibliography{21HgLi}% Produces the bibliography via BibTeX.

\end{document}